\documentclass[lettersize,journal]{IEEEtran}
\usepackage{amsmath,amsfonts}
\usepackage{algorithmic}
\usepackage{algorithm}
\usepackage{array}
\usepackage[caption=false,font=normalsize,labelfont=sf,textfont=sf]{subfig}
\usepackage{textcomp}
\usepackage{stfloats}
\usepackage{url}
\usepackage{verbatim}
\usepackage{graphicx}
\usepackage{cite}
\usepackage{siunitx}
\usepackage[hidelinks]{hyperref}
\usepackage{tuda-pgfplots}
\usepackage{xcolor}
\usepackage{tikz}
\usepackage{tikz-cd}
\usepackage{multirow}

\pgfplotsset{compat=newest}
\usetikzlibrary{%
	calc,
  external,
}
\begin{document}

\title{Foil Conductor Model for Efficient Simulation of {HTS} Coils in Large Scale Applications}

\author{Elias Paakkunainen, Louis Denis, Christophe Geuzaine, Paavo Rasilo, and Sebastian Schöps
\thanks{The work of Elias Paakkunainen is supported by the Graduate School CE within Computational Engineering at the Technical University of Darmstadt. Louis Denis is a research fellow funded by the F.R.S-FNRS. \it{(Corresponding author: Elias Paakkunainen.)}}%
\thanks{Elias Paakkunainen is with the Technical University of Darmstadt, 64289 Darmstadt, Germany, and also with Tampere University, 33720 Tampere, Finland (e-mail: elias.paakkunainen@tu-darmstadt.de).}
\thanks{Louis Denis and Christophe Geuzaine are with the University of Liège, 4000 Liège, Belgium.}
\thanks{Paavo Rasilo is with Tampere University, 33720 Tampere, Finland.}
\thanks{Sebastian Schöps is with the Technical University of Darmstadt, 64289 Darmstadt, Germany.}} 

\markboth{}%
{Paakkunainen \MakeLowercase{\textit{et al.}}: Foil Conductor Model for Efficient Simulation of {HTS} Stacks and Coils in Large Scale Applications}

\IEEEpubid{}

\makeatletter
\def\ps@IEEEtitlepagestyle{
  \def\@oddfoot{\mycopyrightnotice}
  \def\@evenfoot{}
}
\def\mycopyrightnotice{
  {\footnotesize
  \begin{minipage}{\textwidth}
    \fbox{\parbox{\textwidth}{
        This work has been submitted to the IEEE for possible publication. Copyright may be transferred without notice, after which this version may no longer be accessible.
        }}
  \end{minipage}
  }
}

\maketitle

\newcommand{\ez}{\vec{e}_{\mathrm{z}}}

\begin{abstract}
  Homogenization techniques are an appealing approach to reduce computational complexity in systems containing coils with large numbers of high temperature superconductor (HTS) tapes. Resolving all the coated conductor layers and turns in coils is often computationally prohibitive. In this paper, we extend the foil conductor model, well-known in normal conducting applications, to applications with insulated HTS coils. To enhance the numerical performance of the model, the conventional formulation based on $A-V$ is extended to $J-A-V$. The model is verified to be suitable for simulations of superconductors and to accelerate the calculations compared to resolving all the individual layers. The performance of both the $A-V$ and $J-A-V$ formulated models is examined, and the $J-A-V$ variant is concluded to be advantageous. 
\end{abstract}


\section{Introduction}

\IEEEPARstart{R}{EBCO}-based high temperature superconductors (HTS) are currently subject to intensive research and expected to enable various next generation applications. The advantageous properties of HTS enable, e.g., magnets exceeding the field strengths of current designs \cite{Hahn_2019aa,Liu_2020aa} and increased efficiency ratings of power engineering devices. Ongoing efforts aim to develop superconducting electrical machines (motors, generators, transformers) \cite{Haran_2017aa, Chow_2023aa}, power cables \cite{Yazdani-Asrami_2022aa} and fault current limiters \cite{Moyzykh_2021aa,Sotelo_2022aa}, to name a few examples.

Efficient simulation methods are required to reduce the costly and time-consuming prototyping. The complexity of application relevant designs prohibits using analytical solutions. Therefore, numerical methods are commonly used, and the finite element method (FEM) has rapidly become the reference modeling tool for HTS applications \cite{Grilli_2014aa}. Other numerical approaches, such as volume integral methods \cite{Rozier_2019ab} or the Minimum Electro-Magnetic
Entropy Production (MEMEP) method \cite{Pardo_2015aa,Pardo_2017aa}, have also been successfully applied. Independently of the discretization method, various formulations have been proposed. The most popular variants are currently the ones based on $\vec{H}$ \cite{Brambilla_2007aa}, $\vec{H}-\phi$ \cite{Lahtinen_2015aa}, $\vec{T}-\vec{A}$ \cite{Zhang_2017ab}, and $\vec{J}-\vec{A}$ \cite{Stenvall_2010ab}. The choice of formulation is particularly relevant for superconductors since it implicitly determines which material law is used, e.g. $E(J)$ vs. $J(E)$ or $B(H)$ vs. $H(B)$, see e.g. \cite{Sirois_2019aa,Dular_2023aa}.

\begin{figure}
    \centering
    \includegraphics{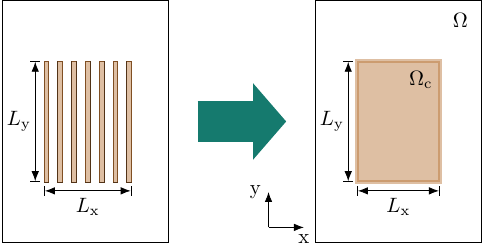}
    \caption{Sketch of a homogenized {HTS} stack. The detailed geometry is replaced with a homogeneous bulk material. Relevant dimensions for the model derivation are displayed.}
    \label{fig:sketch_homogenization}
\end{figure}

Still, the large aspect ratio of the superconducting layers and {HTS} tapes in general, combined with the complexity of the design, often makes a mesh-based discretization of the device under study infeasible in terms of computational time~\cite{Sirois_2015aa}. Several modeling techniques can be used to reduce the complexity and computation time. In particular thin sheet models \cite{Carpenter_1977aa} have been popular for superconducting tapes. They treat the multilayered composite structure as a single (thin) layer \cite{Bortot_2020aa,Schnaubelt_2024aa}. However, a superconducting coil consists of many windings which are still tedious to resolve and eventually require fine meshes leading to unacceptable computational times. This has motivated the application of homogenization on the coil level, see Fig.~\ref{fig:sketch_homogenization}. Geometrical details of the coil cross-section are neglected, and the region is replaced with a homogeneous bulk material having effectively the same electromagnetic behavior as the detailed structure. The homogeneous, i.e. obtained by homogenization, region can be discretized more coarsely than the original detailed geometry.  Models have been presented based on multiple modeling approaches and formulations, e.g. $\vec{H}$ \cite{Zermeno_2013aa}, $\vec{T}-\vec{A}$ \cite{Berrospe-Juarez_2019aa}, and more lately $\vec{J}-\vec{A}$ \cite{Wang_2023aa, Durante-Gomez_2024aa}. In addition, homogeneous models taking into account other than electromagnetic effects, such as thermal effects \cite{Klop_2024aa}, have been presented.

In this paper, we generalize the well-known foil conductor model \cite{De-Gersem_2001aa, Dular_2002aa} to insulated HTS structures. Initially, this model was developed to simulate foil windings made of conventional conducting sheets, typically used in large power inductances and transformers. The topological similarity of foil windings and {HTS} coils motivates the extension of this numerical technique to applications involving many closely arranged {HTS} tapes. Moreover, a change of formulation from $\vec{A}$ to $\vec{J}-\vec{A}$ increases the efficiency of the iterative resolution as it avoids the divergence of the superconducting material constitutive relationship \cite{Dular_2023aa}. The rest of the article is structured as follows. In Section~\ref{sec:formulations}, the model formulations are presented. Section~\ref{sec:verification} contains the verification of the models through numerical experiments on a simple problem, and Section~\ref{sec:example} presents a more advanced problem. Section~\ref{sec:conclusion} concludes the findings of the paper.

\section{Finite element formulations}\label{sec:formulations}
Two formulations of the foil conductor model are presented, namely the $\vec{A}-V-$ and $\vec{J}-\vec{A}-V-$formulations. The voltage $V$ is added to the names to emphasize its importance in the models.
The $\vec{A}-V-$formulation is the direct extension of the normal conducting foil conductor model to applications with {HTS}, while the $\vec{J}-\vec{A}-V-$formulation is a new superconductor-specific suggestion which will be shown to be better suited for {HTS}.

\subsection{The $\vec{A}-V-$formulation}

\begin{figure}
    \centering
    \subfloat[\label{fig:sketch_coil}]{\includegraphics[trim=11em 0em 11em 0em,clip,width=0.23\textwidth]{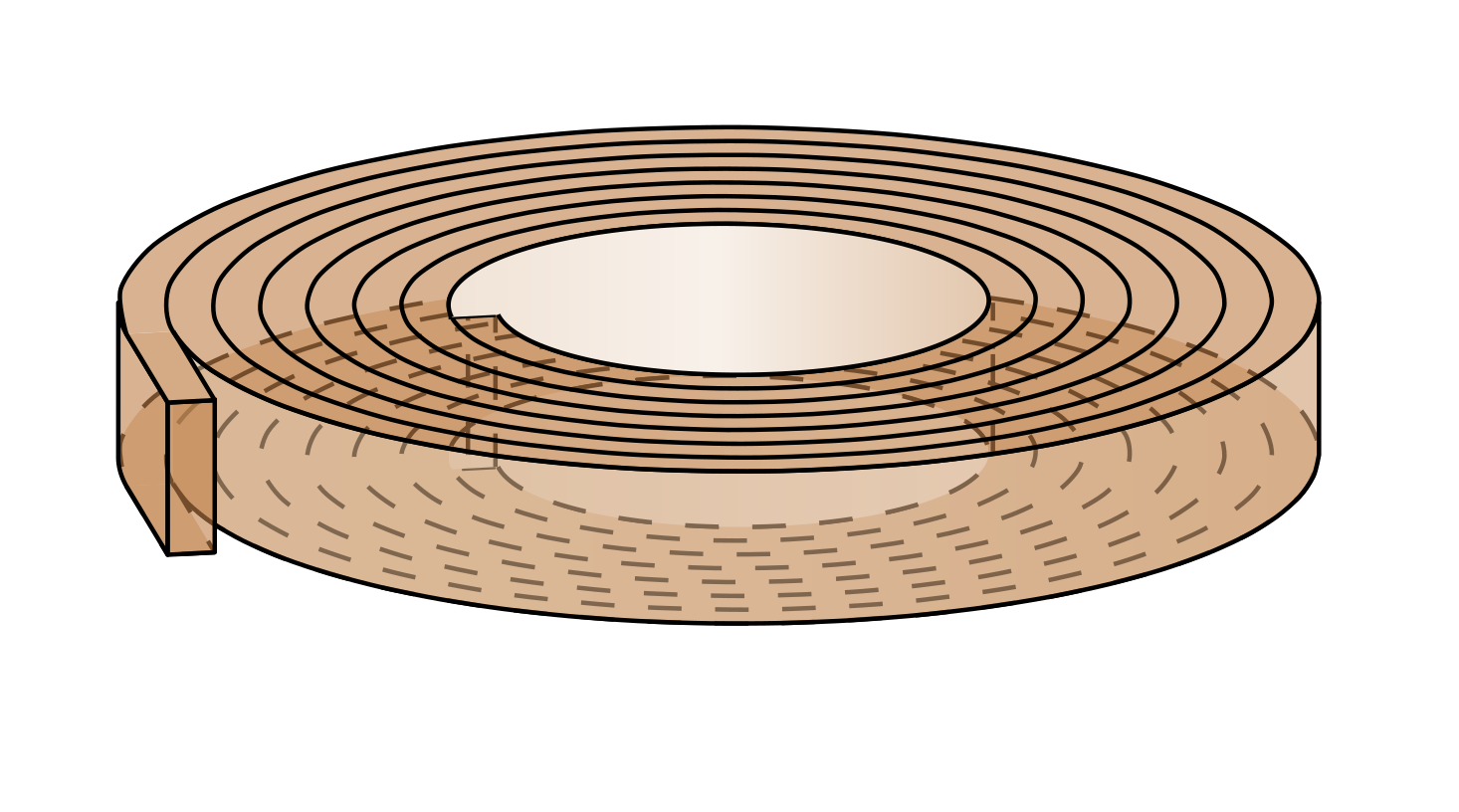}}
    \subfloat[\label{fig:domain_verification}]{\includegraphics[trim=-2em 0em 0em 0em,clip]{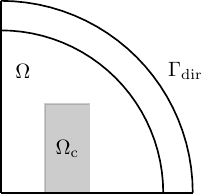}}
    \caption{
        \protect\subref{fig:sketch_coil} Sketch of a coil with a tape stack cross-section.
        \protect\subref{fig:domain_verification} Modeling domain, a quarter of the cross-section of a racetrack coil, which is used for the verification of the homogeneous models.}
    \label{fig:coil_and_domain}
\end{figure}

Let us consider the {2D} domain $\Omega$, sketched in Fig.~\ref{fig:sketch_homogenization} which is the cross-section of a 3D coil, Fig.~\ref{fig:sketch_coil}. In the 2D model, the stack of {HTS} tapes is connected in series and current flow is only considered in the $z$-direction. Note that, the foil conductor model can be written more generally for arbitrary {3D} domains but here we focus on the 2D case. For a more detailed derivation of the model, we refer to \cite{Dular_2002aa} and \cite{Paakkunainen_2024aa}.

We are solving Maxwell's equations in the magnetoquasistatic approximation. In the conducting domain $\Omega_{\mathrm{c}} \subset \Omega$, the electric field $\vec{E}$ can be written in terms of the magnetic vector potential $\vec{A}$ and electric scalar potential $v$ as ${\vec{E} = -\partial_{t}\vec{A} - \text{grad}\,v}$.
Since the current density $\vec{J}$ is assumed to be perpendicular to the {2D} plane, only $z$-components of fields must be considered as functions of $x$ and $y$. In particular, we write $\text{grad}\,v = \Phi\ez$, with $\ez$ being the unit vector in the $z$-direction and $\Phi$ the amplitude of $\text{grad}\,v$. The weak form of the Ampere's law is then written as
\begin{equation}
	\left( \nu_{0}\,\mathrm{curl}\,\vec{A}, \,\mathrm{curl}\,\vec{A}\,^{'} \right)
	+ \left( \sigma\partial_{t}\vec{A}, \,\vec{A}\,^{'} \right)
	+ \left( \sigma\Phi\ez, \,\vec{A}\,^{'} \right)
	= 0,
	\label{eq:fcm_a_field}
\end{equation}
with $\nu_0 = \mu_0^{-1}$ the free space reluctivity ({HTS} tapes are assumed nonmagnetic), $\sigma$ the conductivity and the brackets $(\cdot,\cdot)$ 
denoting the $L^2$ inner product over the domain $\Omega$. The test functions for $\vec{A}$ are denoted as $\vec{A}\,^{'}$. Considering a 2D problem with in-plane magnetic field, $\vec{A}$ can discretized using first order perpendicular edge functions derived from nodal hat functions~\cite{Dular_2020aa}.

Assuming the skin depth to be large in comparison to the thickness of the {HTS} layers, $\vec{J} = \sigma \vec{E}$ is considered to be constant in the $x$-direction in individual layers. In this case, the current density $\vec{J}$ is constrained in each layer as 
\begin{gather}
    \int_{0}^{L_{\mathrm{y}}} \int_{x_i-d/2}^{x_i + d/2} J_{\text{z}}\, dx\,dy= I \quad \Rightarrow \quad \int_{0}^{L_{\mathrm{y}}} J_{\text{z}}\,dy= I/d,
    \label{eq:fcm_a_gen_constraint_louis}
\end{gather}
where $I$ is the net current through each turn, $x_i$ the average $x$-coordinate of turn $i$ and $d$ is the thickness of the coated conductor. From the expression of $\vec{E}$, the current constraint~\eqref{eq:fcm_a_gen_constraint_louis} can be weakly imposed as
\begin{equation}
    \left( \sigma\partial_{t}\vec{A}, \, \Phi^{'}\ez \right) + \left( \sigma\Phi\ez, \, \Phi^{'}\ez \right) + \int_{L_{\mathrm{x}}} \frac{I}{d} \Phi^{'} dx = 0.
    \label{eq:fcm_a_current_constraint_louis}
\end{equation}
Equation \eqref{eq:fcm_a_current_constraint_louis} allows to enforce the current constraint in each layer separately using one single equation.

To ensure a distribution of $\vec{J}$ in $\Omega_{\mathrm{c}}$ which imitates the original layered structure, an additional {1D} discretization depending only on the $x$-coordinate is introduced for $\Phi$
\begin{equation}
    \Phi(x) = \sum_{i=1}^{N_{\mathrm{p}}} u_{i} p_{i}(x),
    \label{eq:phi_discretization}
\end{equation}
where $N_{\mathrm{p}}$ is the number of the basis functions. Multiple choices for the basis functions $p_{i}(x)$ are possible, e.g., functions with local or global support along $L_{\mathrm{x}}$ \cite{Bundschuh_2023af}. Note that the shape functions $p_{i}(x)$ are not required to be defined with respect to the local mesh elements.

The conductivity $\sigma$ of the superconducting material is modeled with the well-known power law expression \cite{Rhyner_1993aa}. To enhance the numerical performance, a regularization term is added to $\sigma$ as described in \cite{Dular_2020aa}, and $\sigma$ is written as
\begin{align}
    \sigma(\|\vec{E}\|) = \frac{J_{\mathrm{c,eng}}}{E_{\mathrm{c}}} \left( \epsilon_{\sigma} + \left( \frac{\|\vec{E}\|}{E_{\mathrm{c}}} \right)^{(n-1)/n} \right)^{-1},
    \label{eq:powerlaw_sigma} 
\end{align}
where $n$ is the power law exponent, $E_{\mathrm{c}}=\SI{e-4}{\volt/\meter}$ the threshold electric field and $\epsilon_{\sigma}$ the small regularization parameter. It should be noted that $\epsilon_{\sigma}$ introduces a small normal conducting component to $\sigma$, which may reduce the accuracy of the numerical results relative to the original power law. The critical current density of the homogeneous material $J_{\mathrm{c,eng}}$ is obtained by scaling the critical current density $J_{\mathrm{c}}$ of the {HTS} material with the fill factor $\lambda$ of the original geometry $J_{\mathrm{c,eng}} = \lambda J_{\mathrm{c}}$, as is commonly done in other homogenizing approaches \cite{Zermeno_2013aa, Berrospe-Juarez_2019aa}. Direct scaling of $J_{\mathrm{c}}$ by $\lambda$ neglects the effect of the normal conducting layers in the {HTS} tapes.

\subsection{The $\vec{J}-\vec{A}-V-$formulation}

The derivation of the $\vec{J}-\vec{A}-V-$formulation follows a similar path, with the exception that in addition to the degrees of freedoms (DoFs) associated with $\vec{A}$ and $\Phi$, the current density $\vec{J}$ also needs to be solved for. Accordingly, new unknowns are added to the system of equations. By solving for $\vec{J}$, the power law can be expressed through the material resistivity $\rho=\sigma^{-1}$ which is expected to lead to improved numerical performance \cite{Dular_2020aa}. The power law resistivity is written as
\begin{equation}
    \rho(\|\vec{J}\|) = \frac{E_{\mathrm{c}}}{J_{\mathrm{c,eng}}} \left( \frac{\|\vec{J}\|}{J_{\mathrm{c}}} \right)^{n-1},
    \label{eq:powerlaw_rho} 
\end{equation}
which contrary to \eqref{eq:powerlaw_sigma} has no added regularization. In this setting, the weak form of the Ampere's law reads
\begin{equation}
	\left( \nu_{0}\,\mathrm{curl}\,\vec{A}, \,\mathrm{curl}\,\vec{A}\,^{'} \right)
	- \left(\vec{J}, \,\vec{A}\,^{'} \right) 
	= 0,
	\label{eq:fcm_ja_field}
\end{equation}
and the weakly imposed current constraint becomes
\begin{equation}
	\left(\vec{J}, \Phi^{'}\ez\,\right)
	- \int_{L_{\mathrm{x}}} \frac{I}{d} \Phi^{'}\,dx = 0.
	\label{eq:fcm_ja_current_constraint}
\end{equation}
These equations were obtained from their counterparts~\eqref{eq:fcm_a_field} and~\eqref{eq:fcm_a_current_constraint_louis} with a substitution of the material property $\vec{J}=\sigma\vec{E}$. The weak form of $\vec{E} = \rho \vec{J}$ is taken as the third equation in the system of equations:
\begin{equation}
	\left( \rho\vec{J}, \, \vec{J}\,^{'} \right) 
	+ \left( \partial_{t} \vec{A}, \, \vec{J}\,^{'} \right)
	+ \left( \Phi\ez, \, \vec{J}\,^{'} \right)
	= 0.
\end{equation}

Since the $\vec{J}-\vec{A}(-V)-$formulation is a mixed formulation, particular care should be taken for its discretization to avoid spurious oscillations \cite{Dular_2021aa}. Following \cite{Wang_2023aa}, we thus discretize $\vec{A}$ with enriched second order perpendicular edge functions and $\vec{J}$ with piecewise constant facet functions. In $\Omega_{\mathrm{c}}^{\mathrm{C}} = \Omega \setminus \Omega_{\mathrm{c}}$, one can restrict $\vec{A}$ to first order perpendicular edge functions to reduce the number of degrees of freedom \cite{Dos-Santos_2024aa}.

\section{Numerical verification}\label{sec:verification}
The homogeneous models described in Section~\ref{sec:formulations} are verified through numerical experiments. The models are compared to a reference model which resolves all the {HTS} layers of the coil and is implemented with the $\vec{A}-$formulation. In the following, as a reference model, we refer to a very finely meshed resolved model which sets the baseline for the comparisons. The resolved model refers to a more coarsely meshed solution that ensures fair comparisons in terms of computation times. All the models were implemented with the open source software GetDP \cite{Dular_1998ac} and Gmsh \cite{Geuzaine_2009ab}. The models are available online in the Life-HTS toolkit\footnote{Available: \href{https://www.life-hts.uliege.be}{www.life-hts.uliege.be}}.

For verification, let us examine the {2D} cross-section of a single racetrack coil in a Cartesian coordinate system. The model domain taking into account the symmetry of the problem is sketched in Fig.~\ref{fig:domain_verification}. A homogeneous Dirichlet boundary condition is set on the outer radius of $\Omega$, and a shell transformation according to \cite{Henrotte_1999aa} is used to reduce the radius of the air domain which has to be discretized. The simulation parameters are listed in Table~\ref{tab:parameters}.
\begin{table}
    \caption{Simulation and material parameters \label{tab:parameters}}
    \centering
    \renewcommand{\arraystretch}{1.3} 
    \begin{tabular}{|c|c|c|}
    \hline Quantity & Symbol & Value \\
    \hline Number of turns & $N$ & \SI{20}{} \\
    \hline Fill factor of {HTS} & $\lambda$ & \SI{0.01}{} \\
    \hline Critical current density & $J_{\mathrm{c}}$ & \SI{e10}{\ampere\per\meter^{2}} \\
    \hline Power law exponent & $n$ & \SI{25}{} \\
    \hline Thickness of {HTS} layer & - & \SI{1}{\micro\meter} \\
    \hline Thickness of the coated conductor & $d$ & \SI{100}{\micro\meter} \\
    \hline Coated conductor width & - & \SI{12}{\milli\meter} \\
    \hline Frequency & $f$ & \SI{50}{\hertz} \\
    \hline 
    \end{tabular}
\end{table}

\begin{figure}
    \centering
    \includegraphics{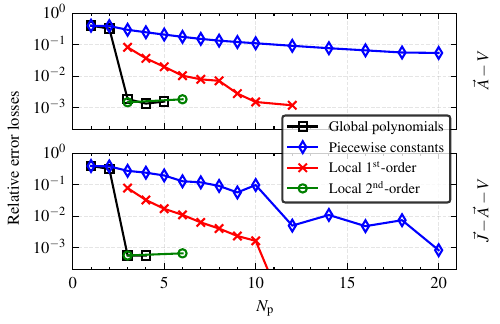}
    \caption{Relative error of the losses $p$ predicted with $\vec{A}-V$ (top) and $\vec{J}-\vec{A}-V$ (bottom) homogeneous models compared to the reference model. Good accuracies are obtained with different basis function choices for $\Phi$, given that the number of basis functions $N_{\mathrm{p}}$ is sufficiently large.}
    \label{fig:verification_relerr}
\end{figure}

\begin{figure}
    \centering
    \includegraphics{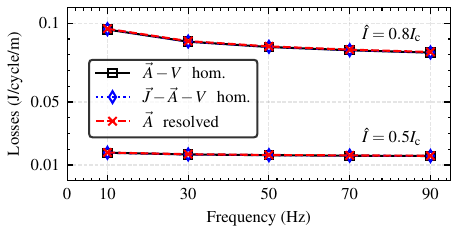}
    \caption{Comparison of the losses $p$ obtained with the $\vec{J}-\vec{A}-V$ homogeneous, $\vec{A}-V$ homogeneous and resolved models as a function of frequency for two different current amplitudes.}
    \label{fig:verification_losses}
\end{figure}

First, we examine the accuracy of the homogeneous models by comparing the predicted losses $p$ to the losses predicted with the reference model. A current is imposed through the coil with an amplitude of $\hat{I}=0.8I_{\mathrm{c}}$. Fig.~\ref{fig:verification_relerr} shows the relative error of the losses with different choices and numbers of basis functions for $\Phi$. The relative error is defined as $\left| p - p_{\mathrm{ref}} \right| / p_{\mathrm{ref}}$.
A good accuracy is obtained in all cases with a sufficient amount of basis functions except for the $\vec{A}-V$ homogeneous model with the piecewise constants where the convergence is slow. Higher order basis functions are observed to converge faster. For the rest of the numerical tests in this article, a $3^{\text{rd}}$\,-order global polynomial is the standard choice for the discretization of $\Phi$. Fig.~\ref{fig:verification_losses} shows the good agreement in the losses over a range of frequencies for two different current amplitudes.

\begin{figure}
    \centering
    \includegraphics[width=\columnwidth]{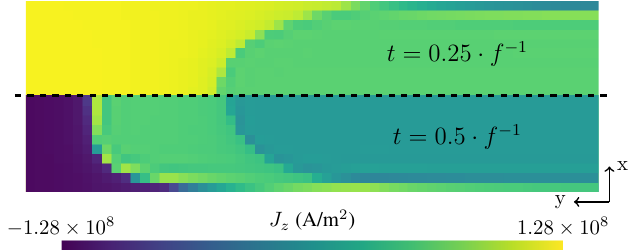}
    \caption{Distributions of $\vec{J}$ in the coil using the $\vec{J}-\vec{A}-V$ homogeneous model when a current with $\hat{I}=0.8I_{\mathrm{c}}$ is imposed. The distribution is shown at the peak $I$ (top) and after one half-cycle (bottom).}
    \label{fig:verification_j_dist}
\end{figure}

The distribution of $\vec{J}$ in $\Omega_{\mathrm{c}}$ obtained with the $\vec{J}-\vec{A}-V$ homogeneous model is shown in Fig.~\ref{fig:verification_j_dist}. Distributions resembling those from coils are obtained in the homogeneous region.
The numerical performance of the models is compared in Table~\ref{tab:sim_times}. The homogeneous models are faster to solve than the resolved model, especially when the number of turns increases. The very finely discretized reference and a resolved solution allowing an error of the same order of magnitude as the homogeneous models are listed separately. The $\vec{J}-\vec{A}-V$ and $\vec{A}-V$ homogeneous models have roughly similar computation times despite the $\vec{J}-\vec{A}-V$ formulated model having significantly more DoFs. The number of DoFs is larger for the $\vec{J}-\vec{A}-V-$formulation because it requires both additional solving for $\vec{J}$ and the enrichment of $\vec{A}$ in $\Omega_{\mathrm{c}}$. The slower rate of convergence for the nonlinear iterations due to using the power law conductivity causes the $\vec{A}-V$ homogeneous model to lose the advantage gained in terms of DoFs. Additionally, $\epsilon_{\sigma}$ is required for convergence and has been observed to be simulation specific. In the simulations carried out in this paper, the value of $\epsilon_{\sigma}$ varies between \SI{e-8}{} and \SI{5e-4}{}. The $\vec{A}-V$ homogeneous model is observed to require a larger $\epsilon_{\sigma}$ value compared to the resolved model.

\begin{table}
    \caption{Verification simulation - comparison of the models\label{tab:sim_times}}
    \centering
    \renewcommand{\arraystretch}{1.55} 
    \begin{tabular}{|c|c|c|c|c|c|}
    \hline 
    $N$ & Model & Formulation & Time & DoFs & Rel. err. \\
    \hline
    \multirow{4}{*}{$20$} & Reference & $\vec{A}$ & \SI{58}{\minute} & \SI{22.8}{k} & - \\ 
    \cline{2-6} & Resolved & $\vec{A}$ & \SI{30}{\minute} & \SI{13.1}{k} & \SI{5.1e-4}{} \\ 
    \cline{2-6} & Homog. & $\vec{A}-V$ & \SI{13}{\minute} & \SI{3.1}{k} & \SI{4.7e-3}{} \\ 
    \cline{2-6} & Homog. & $\vec{J}-\vec{A}-V$ & \SI{9}{\minute} & \SI{8.0}{k} & \SI{5.0e-4}{} \\ 
    \hline
    \hline
    \multirow{4}{*}{$50$} & Reference & $\vec{A}$ & \SI{175}{\minute} & \SI{69.9}{k} & - \\ 
    \cline{2-6} & Resolved & $\vec{A}$ & \SI{61}{\minute} & \SI{24.4}{k} & \SI{1.2e-4}{} \\ 
    \cline{2-6} & Homog. & $\vec{A}-V$ & \SI{21}{\minute} & \SI{5.0}{k} & \SI{2.3e-3}{} \\ 
    \cline{2-6} & Homog. & $\vec{J}-\vec{A}-V$ & \SI{22}{\minute} & \SI{14.7}{k} & \SI{2.6e-3}{} \\ 
    \hline
    \hline
    \multirow{4}{*}{$100$} & Reference & $\vec{A}$ & \SI{838}{\minute} & \SI{299.4}{k} & - \\ 
    \cline{2-6} & Resolved & $\vec{A}$ & \SI{233}{\minute} & \SI{101.5}{k} & \SI{1.3e-3}{} \\ 
    \cline{2-6} & Homog. & $\vec{A}-V$ & \SI{37}{\minute} & \SI{8.3}{k} & \SI{3.4e-3}{} \\ 
    \cline{2-6} & Homog. & $\vec{J}-\vec{A}-V$ & \SI{47}{\minute} & \SI{27.6}{k} & \SI{2.7e-3}{} \\ 
    \hline 
    \end{tabular}
\end{table}

\section{Simulation example}\label{sec:example}
As a more complicated problem, we examine a stack of racetrack coils. Again, the symmetry of the problem is utilized, and the modeling domain is similar to Fig.~\ref{fig:domain_verification}. The parameters are chosen as in Table~\ref{tab:parameters}, except that we now consider 5 racetracks with $N=50$ layers each with \SI{5}{\milli\meter} gaps separating the coils from each other. Again, a current with $\hat{I}=0.8I_{\mathrm{c}}$ is imposed.

\begin{figure}
    \centering
    \includegraphics[width=\columnwidth]{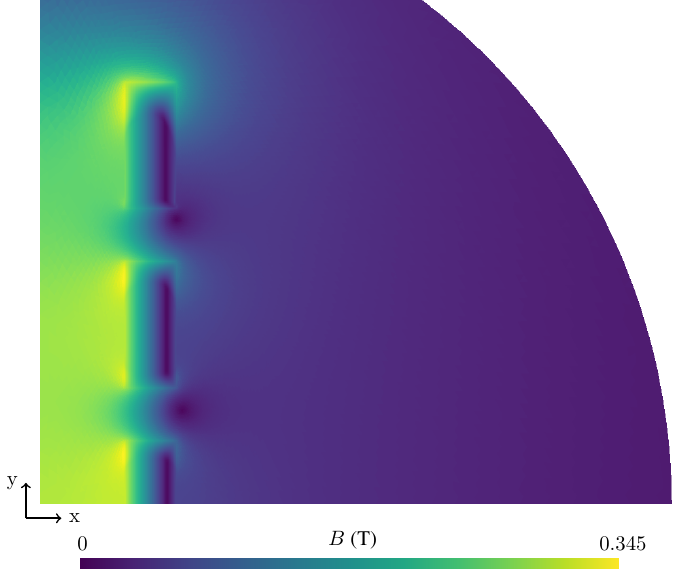}
    \caption{Magnitude of the magnetic flux density $B$ in the stack of racetrack coils obtained with the $\vec{J}-\vec{A}-V$ homogeneous model. The field is plotted at peak current value.}
    \label{fig:sim_example_b_plot}
\end{figure}

Table~\ref{tab:sim_example_results} summarizes the results obtained with the homogeneous models. Similar conclusions can be drawn from the results as in the previous example in Section~\ref{sec:verification}. Both homogeneous models predict similar results in terms of losses, and the computation times remain similar. Again the trade-off between the number of DoFs and rate of convergence is observed. Fig.~\ref{fig:sim_example_b_plot} shows the magnetic flux density in the homogeneous stack of coils.

\begin{table}
    \caption{Simulation example - comparison of the homogeneous models\label{tab:sim_example_results}}
    \centering
    \renewcommand{\arraystretch}{1.55} 
    \begin{tabular}{|c|c|c|c|c|}
    \hline  Model & Time (min.) & DoFs & Losses (J/cycle/m)  \\
    \hline  $\vec{A}-V$ hom. & \SI{72}{} & \SI{27.3}{k} & \SI{2.16}{} \\
    \hline  $\vec{J}-\vec{A}-V$ hom. & \SI{81}{} & \SI{57.4}{k} & \SI{2.15}{} \\
    \hline 
    \end{tabular}
\end{table}

\section{Conclusion}\label{sec:conclusion}
The foil conductor model has been shown to be suitable for the simulation of HTS coils. Both the $\vec{A}-V$ and ${\vec{J}-\vec{A}-V}$ homogeneous models are verified to be accurate, and to provide a significant speedup compared to resolving all the layers. The $\vec{J}-\vec{A}-V$ formulated model is concluded to have better numerical properties despite having higher number of DoFs due to the additional unknowns and the function space enrichment. The need for added regularization and the slow convergence of the nonlinear solver make the $\vec{A}-V$ homogeneous model a less appealing choice. In contrary, the $\vec{J}-\vec{A}-V$ homogeneous model demonstrated robust convergence in our numerical experiments without the need to find simulation specific regularization values. The $\vec{J}-\vec{A}-V$ homogeneous model is a promising approach for the simulation of HTS coils, particularly in 2D.



\begin{thebibliography}{1}
    \bibliographystyle{IEEEtran}

    \bibitem{Hahn_2019aa}
    S. Hahn \emph{et al.}, “45.5-tesla direct-current magnetic field generated with a high-temperature superconducting magnet,” \emph{Nature}, vol. 570, no. 7762, pp. 496-499, 2019.

    \bibitem{Liu_2020aa}
    J. Liu \emph{et al.}, “World record 32.35 tesla direct-current magnetic field generated with an all-superconducting magnet,” \emph{SUST}, vol. 33, no. 3, p. 03LT01, Feb. 2020.

    \bibitem{Haran_2017aa} K. S. Haran et al., “High power density superconducting rotating machines - development status and technology roadmap,” \emph{SUST}, vol. 30, no. 12, p. 123002, Dec. 2017.
    
    \bibitem{Chow_2023aa} C. C. Chow, M. D. Ainslie, and K. Chau, “High temperature superconducting rotating electrical machines: an overview,” \emph{Energy Reports}, vol. 9, pp. 1124-1156, Dec. 2023.
    
    \bibitem{Yazdani-Asrami_2022aa} M. Yazdani-Asrami, S. Seyyedbarzegar, A. Sadeghi, W. T. B. de Sousa, and D. Kottonau, “High temperature superconducting cables and their performance against short circuit faults: current development, challenges, solutions, and future trends,” \emph{SUST}, vol. 35, no. 8, p. 083002, Jul. 2022.

    \bibitem{Moyzykh_2021aa}
    M. Moyzykh \emph{et al.}, “First russian 220 kV superconducting fault current limiter {(SFCL)} for application in city grid,” \emph{{IEEE} Trans. Appl. Super.}, vol. 31, no. 5, pp. 1-7, 2021.

    \bibitem{Sotelo_2022aa}
    G. G. Sotelo \emph{et al.}, “A review of superconducting fault current limiters compared with other proven technologies,” \emph{Superconductivity}, vol. 3, p. 100018, 2022.

    \bibitem{Grilli_2014aa}
    F. Grilli, E. Pardo, A. Stenvall, D. N. Nguyen, W. Yuan, and F. Gomory, “Computation of losses in HTS under the action of varying magnetic fields and currents,” \emph{{IEEE} Trans. Appl. Super.}, vol. 24, no. 1, pp. 8200333, Feb. 2014.

    \bibitem{Rozier_2019ab}
    B. Rozier, A. Badel, B. Ramdane, and G. Meunier, “Calculation of the local current density in high-temperature superconducting insulated rare earth-barium-copper oxide coils using a volume integral formulation and its contribution to coil protection,” \emph{SUST}, vol. 32, no. 4, p. 044008, Mar. 2019.

    \bibitem{Pardo_2015aa}
    E. Pardo, J. Souc, and L. Frolek, “Electromagnetic modelling of superconductors with a smooth current-voltage relation: variational principle and coils from a few turns to large magnets,” \emph{SUST}, vol. 28, no. 4, p. 044003, Feb. 2015.

    \bibitem{Pardo_2017aa}
    E. Pardo and M. Kapolka, “{3D} computation of non-linear eddy currents: Variational method and superconducting cubic bulk,” \emph{J. Comput. Phys.}, vol. 344, pp. 339–363, 2017.
    
    \bibitem{Brambilla_2007aa}
    R. Brambilla, F. Grilli, and L. Martini, “Development of an edge-element model for {AC} loss computation of high-temperature superconductors,” \textit{SUST}, vol. 20, no. 1, pp. 16--24, Jan. 2007.

    \bibitem{Lahtinen_2015aa}
    V. Lahtinen, A. Stenvall, F. Sirois, and M. Pellikka, “A finite element simulation tool for predicting hysteresis losses in superconductors using an H-oriented formulation with cohomology basis functions,” \textit{J. Supercond. Nov. Magn.}, vol. 28, no. 8, pp. 2345--2354, Aug. 2015.

    \bibitem{Zhang_2017ab}
    H. Zhang, M. Zhang, and W. Yuan, “An efficient 3D finite element method model based on the {T-A} formulation for superconducting coated conductors,” \emph{SUST}, vol. 30, no. 2, p. 024005, Feb. 2017.

    \bibitem{Stenvall_2010ab}
    A. Stenvall and T. Tarhasaari, “An eddy current vector potential formulation for estimating hysteresis losses of superconductors with FEM,” \emph{SUST}, vol. 23, no. 12, p. 125013, Nov. 2010.

    \bibitem{Sirois_2019aa} F. Sirois, F. Grilli, and A. Morandi, “Comparison of constitutive laws for modeling high-temperature superconductors,” \emph{{IEEE} Trans. Appl. Super.}, vol. 29, no. 1, p. 8000110, Jan. 2019.

    \bibitem{Dular_2023aa} J. Dular, “Standard and mixed finite element formulations for systems with type-II superconductors,” PhD Thesis, Université de Liège, 2023.

    \bibitem{Sirois_2015aa}
    F. Sirois and F. Grilli, “Potential and limits of numerical modelling for supporting the development of HTS devices,” \textit{SUST}, vol. 28, no. 4, p. 043002, Apr. 2015.

    \bibitem{Carpenter_1977aa} C. J. Carpenter, “Comparison of alternative formulations of 3-dimensional magnetic-field and eddy-current problems at power frequencies,” \emph{Proceedings of the Institution of Electrical Engineers}, vol. 124, no. 11, pp. 1026--1034, 1977.

    \bibitem{Bortot_2020aa} L. Bortot \emph{et al.}, “A coupled A-H formulation for magneto-thermal transients in high-temperature superconducting magnets,” \emph{{IEEE} Trans. Appl. Super.}, vol. 30, no. 5, Aug. 2020.

    \bibitem{Schnaubelt_2024aa} E. Schnaubelt \emph{et al.}, “Magneto-thermal thin shell approximation for {3D} finite element analysis of no-insulation coils,”  \emph{{IEEE} Trans. Appl. Super.}, vol. 34, no. 3, pp. 1--6, May 2024.

    \bibitem{Zermeno_2013aa} V. M. R. Zermeno, A. B. Abrahamsen, N. Mijatovic, B. B. Jensen, and M. P. Sørensen, “Calculation of alternating current losses in stacks and coils made of second generation high temperature superconducting tapes for large scale applications,” \emph{JAP}, vol. 114, no. 17, p. 173901, Nov. 2013.
    
    \bibitem{Berrospe-Juarez_2019aa} E. Berrospe-Juarez, V. M. R. Zermeno, F. Trillaud, and F. Grilli, “Real-time simulation of large-scale HTS systems: multi-scale and homogeneous models using the T-A formulation,” \emph{SUST}, vol. 32, no. 6, p. 065003, Apr. 2019.

    \bibitem{Wang_2023aa} S. Wang, H. Yong, and Y. Zhou, “Numerical calculations of high temperature superconductors with the {J-A} formulation,” \emph{SUST}, vol. 36, no. 11, p. 115020, Sep. 2023.
        
    \bibitem{Durante-Gomez_2024aa} W. Durante-Gomez \emph{et al.}, “Fem-circuit co-simulation of superconducting synchronous wind generators connected to a {DC} network using the homogenized {J-A} formulation of the {Maxwell} equations,” \emph{SUST}, vol. 37, no. 6, p. 065021, May 2024.
    
    \bibitem{Klop_2024aa} C. L. Klop, R. Mellerud, C. Hartmann, and J. K. Noland, “Electro-thermal homogenization of {HTS} stacks and {Roebel} cables for machine applications,” \emph{{IEEE} Trans. Appl. Super.}, vol. 34, no. 4, pp. 1-10, Jun. 2024.

    \bibitem{De-Gersem_2001aa} H. De Gersem and K. Hameyer, “A finite element model for foil winding simulation,” \emph{{IEEE} Trans. Magn.}, vol. 37, no. 5, pp. 3427-3432, Sep. 2001.

    \bibitem{Dular_2002aa} P. Dular and C. Geuzaine, “Spatially dependent global quantities associated with {2-D} and {3-D} magnetic vector potential formulations for foil winding modeling,” \emph{{IEEE} Trans. Magn.}, vol. 38, no. 2, pp. 633-636, Mar. 2002.

    \bibitem{Paakkunainen_2024aa} E. Paakkunainen, J. Bundschuh, I. C. Garcia, H. De Gersem, and S. Schöps, “A stabilized circuit-consistent foil conductor model,” \emph{{IEEE} Access}, vol. 12, pp. 1408-1417, 2024.

    \bibitem{Dular_2020aa} J. Dular, C. Geuzaine, and B. Vanderheyden, “Finite-element formulations for systems with high-temperature superconductors,” \emph{{IEEE} Trans. Appl. Super.}, vol. 30, no. 3, pp. 1-13, Apr. 2020.

    \bibitem{Bundschuh_2023af} J. Bundschuh, Y. Späck-Leigsnering, and H. De Gersem, “Homogenization of foil windings with globally supported polynomial shape functions,” \emph{Arch. Electr. Eng.}, vol. 73, no. 1, pp. 77–85, 2024.

    \bibitem{Rhyner_1993aa}
    J. Rhyner, “Magnetic properties and {AC}-losses of superconductors with power law current—voltage characteristics,” \emph{Physica C: Superconductivity,} vol. 212, no. 3-4, pp. 292-300, 1993.

    \bibitem{Dular_2021aa} J. Dular, M. Harutyunyan, L. Bortot, S. Schöps, B. Vanderheyden, and C. Geuzaine, “On the stability of mixed finite-element formulations for high-temperature superconductors,” \emph{{IEEE} Trans. Appl. Super.}, vol. 31, no. 6, Sep. 2021.

    \bibitem{Dos-Santos_2024aa}
    G. dos Santos and B. M. Oliveira, “A shape function order approach to accelerate the computation time of the {J-A} formulation,” \emph{TechRxiv}, 2024.

    \bibitem{Dular_1998ac} P. Dular, C. Geuzaine, F. Henrotte, and W. Legros, “A general environment for the treatment of discrete problems and its application to the finite element method,” \emph{{IEEE} Trans. Magn.}, vol. 34, no. 5, pp. 3395-3398, Sep. 1998.

    \bibitem{Geuzaine_2009ab} C. Geuzaine and J.-F. Remacle, “Gmsh: A {3-D} finite element mesh generator with built-in pre- and post-processing facilities,” \emph{Int. J. Numer. Meth. Eng.}, vol. 79, pp. 1309-1331, 2009.
    
    \bibitem{Henrotte_1999aa} F. Henrotte, B. Meys, H. Hedia, P. Dular, and W. Legros, “Finite element modelling with transformation techniques,” \emph{{IEEE} Trans. Magn.}, vol. 35, no. 3, pp. 1434-1437, May 1999.

\end{thebibliography}
\end{document}